\documentstyle[aps]{revtex}
\begin{document}
\twocolumn[
\date{\today} 
\title
{Neutral Meson Photoproduction in  $SU_f(3)$  $\chi$PT \\
(I): $\gamma N \rightarrow \pi^0 N$}
\author{M. K. Banerjee and J. Milana}
\address
{Department of Physics, University of Maryland \\
College Park, Maryland 20742, U.S.A.}
\date{DOE/ER/40762--077, UMPP \#96--062,
nucl-th/9601036, January 24, 1996} 
\maketitle

\begin{abstract}\widetext
We present the results for the electric dipole amplitude for 
$\gamma N \rightarrow \pi^0 N$ at threshold at the $O(p^2)$ 
level in SU$_f$(3) chiral perturbation theory.    We find that the
SU$_f$(3) results 
differ only slightly from   the SU$_f$(2) results.   
At  the $O(p^3)$ level one encounters new, unknown counterterms to fix
 which one is likely to need the threshold photoproduction data
themselves, thus losing predictive power.   We suggest, instead,  that the {\it
difference} between the proton and 
neutron $\pi^0$ photoproduction amplitudes may  provide a   test of the
convergence properties of the $\chi$PT in the present context.  We urge that the
neutron's electric dipole amplitude 
be measured.
\end{abstract}
\vglue0.25in ]

\narrowtext
\section{Introduction}
The one--loop $O(p^2)$ contributions in SU$_f(2)$ chiral perturbation
theory ($\chi$PT) 
 to the photoproduction rates at threshold of neutral pions from nucleons are 
known\cite{andGasser} to lead to a large, positive contribution.  
Experimentally however\cite{pidata}  the electric dipole amplitude $E_{0^+}$ 
is found to be negative and fairly well described by tree--level 
contributions\cite{AaronBarry}.   This situation is 
somewhat of an embarrassment for $\chi$PT.  Bernard, Kaiser and Mei{\ss}ner 
have argued that the  $E_{0^+}$ has a slowly convergent, perturbative expansion,  
 having recently calculated\cite{latestBernard} the one--loop  $O(p^3)$  
contributions.   However at this order ultraviolet divergences appear in the 
loop evaluations which are renormalized by the further contribution of  unknown 
``counterterms'' that enter   the chiral Lagrangian at this order (one each for 
the proton and the neutron, reflecting the separate isoscalar and
isovector coupling 
of the photon). Hence formally there is no prediction from  $\chi$PT for 
    the   electric   dipole   amplitudes   at   this    
order.\footnote{In the work\cite{latestBernard} of Bernard {\it et al.}, 
a model employing resonance saturation was used to estimate the
residual finite pieces of the unknown  ``counterterms'', 
from which they thus were able to fit the data 
reasonably well.}   

The work of Refs.\cite{andGasser} and \cite{latestBernard} were within the 
framework of SU$_f$(2) $\chi$PT.  
In the present work we explore the results of including strangeness.  We
restrict ourselves to 
only the one--loop $O(p^2)$ level, where $\chi$PT makes unambiguous predictions for 
the electric dipole amplitude (at threshold).  The motivation for this
analysis is the 
observation that in other electromagnetic processes that have been
studied at the one--loop 
 level in $\chi$PT\cite{gangof4,us2}, there is a tendency for the kaon and 
pion loops to    cancel partially.   As we will see however, this  does not occur 
in the present case.
We believe that  whether the chiral expansion 
is in fact converging either in the SU$_f$(2) sector or in the
SU$_f(3)$ sector is an open question.  

We suggest  a test of convergence of     based on the
predicted difference between the $\gamma+p\rightarrow  \pi^0 + p$ and
(as yet unmeasured) $\gamma+n\rightarrow  \pi^0 + n$  amplitudes.
Unlike the individual amplitudes   this difference 
has a well behaved perturbative expansion   at
the one--loop $O(p^2)$ level. 

An important component of the strangeness program at CEBAF involves kaon and eta 
meson production.  As in the case of $\pi^0$ photoproduction, at  $O(p^2)$, 
SU$_f$(3)chiral perturbation theory makes  
predictions for threshold $K^0$ and $\eta$ photoproduction rates 
(although once again the issue of convergence arises).    
These rates however require a special handling of resonances which 
are known\cite{known1,known2} to play an 
important role.\footnote{The relevant resonances are 1/2$^-$ states such as the 
$S_{11}(1535)$ in the case of $\eta$ photoproduction.  The 3/2$^+$ states such as 
the $\Delta(1232)$ first enter at tree--level at $O(p^2)$.   Their
contribution is though numerically suppressed relative to 
the nucleon's anomalous magnetic moment terms.    Since the main issue
involves the $O(p^2)$ one--loop contributions, we will here follow 
Ref. \cite{andGasser} and keep only the nucleon  explicit.}  
Because the $\pi^0$ photoproduction amplitudes are thus in many ways 
cleaner and already involve distinct theoretical issues that can be
precisely formulated, and because of the experimental interest in the subject, 
we have chosen to present 
here the   results for pion production amplitudes only and will
defer addressing the strange 
meson production to a subsequent paper. 

The remainder of this paper is organized in the following manner.  In
section (II) we
review the tree--level 
contributions, emphasizing the role\cite{Bernardetal} that the relativistic
theory plays in $\chi$PT in 
determining the $O(p^2)$ amplitudes.  In section (III) we present the
one--loop, $O(p^2)$ 
results as well as the full expression for the electric dipole
amplitudes at this order in 
$SU_f(3)$ $\chi$PT.  In section (IV) we discuss these results and their
theoretical and 
experimental implications.  We also describe our proposed test of
convergence of the   theory using the amplitude difference 
$E_{0^+}\left(\gamma+p\rightarrow\pi^0+p\right) - 
E_{0^+}\left(\gamma+n\rightarrow\pi^0+n\right)$.  Section (V)
contains our conclusions and outlooks.

\section{Tree level contributions}

The amplitude for threshold neutral meson production can be written as
\cite{Bernardetal} 
\begin{equation}
{\cal M}   = \imath\bar{H} S \cdot \epsilon \, C_1 \, H,
\label{Mdef}
\end{equation}
in terms of which $E_{0^+}$ is defined as
\begin{equation}
E_{0^+} = - \frac{C_1}{8 \pi (1 + m/M_N)},
\label{dipoledef}
\end{equation}
where $m$ is the produced meson's mass, $M_N$ the    nucleon mass,  
$\epsilon_\mu$ is the photon  polarization vector and $S^\mu_v \equiv \frac12
\gamma_5(\gamma_\mu-v\!\!\!/v_\mu)$ is the transverse  spin operator  
defined in Ref.\cite{JenkMan1} and in which $v$ is the nucleon's four velocity.    
Note that a Taylor expansion of 
the above denominator is required in the chiral expansion of $E_{0^+}$.  
   The tree--level contributions through $O(p^2)$ are 
determined\cite{AaronBarry} from the relativistic  theory.   These are shown 
in Fig. (1).  The  $O(p^1)$ term is evaluated using the lowest--order terms 
of the relativistic Lagrangian: 
\begin{eqnarray}
{\cal L}_{eff} &=& {\cal L}_0^{\pi N} + {\cal L}_2^{\pi \pi}
\nonumber\\
{\cal L}_0^{\pi N} &=& Tr \overline{B} (i\not\!\!D - M_B) B +
\nonumber\\
&&D Tr \overline{B} \gamma^\mu \gamma_5 \{A_\mu, B\} + F Tr
\overline{B} \gamma^\mu \gamma_5 [A_\mu,B] \nonumber\\
{\cal L}_2^{\pi \pi} &=& \frac{f_\pi^2}{4} Tr \partial_\mu \Sigma
\partial^\mu \Sigma^\dagger
+ a Tr M(\Sigma + \Sigma^\dagger),
\label{lagfull}
\end{eqnarray}   
\noindent in which, $\xi = e^{i \pi/f_\pi}$, $\Sigma = \xi^2 = e^{i
2\pi/f_\pi}$, and
\begin{eqnarray}
&V_\mu &= \frac{1}{2}
[\xi\partial_\mu\xi ^\dagger + \xi^\dagger
(\partial_\mu \xi)],\\ 
&A_\mu &= \frac{i}{2}[\xi\partial_\mu\xi ^\dagger - \xi^\dagger
(\partial_\mu \xi)]  ,\nonumber\\
&D^\mu B &= \partial^\mu B + [V^\mu, B].\hspace{.3in}
\end{eqnarray}
The definitions of the mass matrix, $M$, and the octet meson and
baryon fields are, by now, standard, and can be found, for example, 
in Ref.~\cite{pedagogy}.   

The $O(p^2)$ tree--level  contribution is a combination of two terms.   
The first is from the aforementioned Taylor expansion of the denominator in 
Eq.~(\ref{dipoledef}) with the $O(p^1)$ numerator.  The second is from the 
  photon's anomalous magnetic moment couplings\cite{ColeGlas} with the nucleon in 
Figs. (1a) and (1b), 
\begin{eqnarray}
{\cal L}_1^{\gamma N} = \frac{e}{4 M_N}&&\left( 
\kappa_D Tr \overline{B} \sigma_{\mu \nu}F^{\mu \nu}\{ Q, B \} \right. \nonumber\\
&&\left. + \kappa_F Tr \overline{B} \sigma_{\mu \nu}F^{\mu \nu} [ Q, B] \right).
\end{eqnarray}
$Q$ in the above is the charge matrix of the $u, d$ and $s$ quarks.  The 
coefficients $\kappa_D$ and $\kappa_F$ must be determined from data, to wit 
one can use as input the measured anomalous magnetic moments 
of the proton and neutron in terms of which 
$\kappa_D = - \frac32 \kappa_n \approx 2.87$ and 
$\kappa_F = \kappa_p  + \frac12 \kappa_n \approx .84$.

In the heavy--baryon expansion of $\chi$PT\cite{JenkMan1} 
in which all negative energy states of the nucleon are absorbed into 
higher--order terms of the chiral expansion, 
${\cal L}_0^{\pi N}$ above becomes  
\begin{eqnarray}
{\cal L}_{v}^0 &=& i Tr \overline{B}_v v\cdot D B_v +
\nonumber\\
&&2 D Tr \overline{B_v} S^\mu_v \{A_\mu, B_v\} +2 F Tr
\overline{B_v} S^\mu_v [A_\mu, B_v]. \label{Lheavy}
\end{eqnarray}        
The nucleon's propagator is given directly to be $i/(v\cdot k + i\eta)$.

The $O(p^1)$ tree--level contribution to neutral pion photoproduction 
 in the heavy baryon expansion arises from the following term in 
${\cal L}^1_{v}$\cite{JenkMan1}, 
\begin{equation}
f_2 Tr \overline{B}_v S_v \cdot D v \cdot A B_v \,\in \,{\cal L}^1_{v}.
\end{equation}
Bernard {\it et al}~\cite{Bernardetal} determined the coefficient $f_2$
 by matching with the amplitude obtained in the 
 relativistic theory.  The legitimacy of the procedure relies 
on the fact\cite{andGasser} that in the relativistic theory acceptable
``counterterms''  
({\it i.e.} that are Hermitian and CPT invariant)  enter $E_{0^+}$  first
at $O(p^3)$ only and not at a 
lower level.\footnote{This last point
may not be obvious at first. 
For example one may  consider adding the following CPT invariant,
Hermitian pair of terms to the 
relativistic theory:
\begin{equation}
\frac{g_2}{2 M_N} \overline{B}\left[ -i \not\!\!\loarrow{D} \gamma_5 \not\!\!A 
+ \gamma_5 \not\!\!A i \not\!\!\roarrow{D} \right] B.
\end{equation}
However such terms are completely absorbed by a chiral rotation of the
nucleon field 
$B^\prime = ( 1 + \frac{g_2}{2 M_N} \gamma_5 \not\!\!A) B$.   
Explicit evaluation also shows that the graphs of Fig. (1) are independent of $g_2$ 
provided that the axial Noether current  has been constructed including
the effect of the new interaction term.} 
 
The fact that $f_2$ is determined from the relativistic theory is 
worth emphasizing.    If this was not the case and $f_2$ was a free parameter, 
then,  in fact, there would be no prediction from chiral perturbation 
theory of pion photoproduction as the latter would be the 
source of fixing this now free parameter.  However this is not the case because  
the heavy baryon expansion is ultimately merely a reorganization 
(albeit an admittedly highly convenient and often more transparent one) 
of the full relativistic theory.    In the
relativistic theory the loop expansion does not correspond in a one to one 
fashion with the chiral expansion\cite{GassNuc} as it does in   
Weinberg's power counting method\cite{Wcount}. Instead,  because the
nucleon's mass is explicit in the theory,  it generates  terms   at
lower orders in the chiral expansion as well as $m/M_N$ corrections 
to all order.    This fact however  must ultimately be  said to be only a 
complicating feature of the relativistic formulation.    
The key feature is that the chiral power of the $M_N$ independent term 
generated by the loop expansion in the relativistic theory is the same as in the 
heavy baryon expansion.     As in any perturbative expansion, 
the coefficients of the lower order terms always must be readjusted when going to 
higher order in the expansion and hence whether they formally contain
pieces proportional to $M_N$ is irrelevant.    The  role of any term
$\sim (m/M_N)^n$   generated   from the relativistic  loop expansion is
included in a heavy baryon expansion provided all permissible terms of
the specified chiral power are included.   Therefore, while the
explicit elimination of the nucleon's mass 
is a greatly simplifying feature of the heavy baryon expansion,  the
relativistic theory is   nevertheless essential.  The determination  of
  $f_2$ by Bernard {\it et al}~\cite{Bernardetal} illustrates this point.  

Having now reviewed their origin, we conclude this section by 
listing the tree--level  contributions through $O(p^2)$ in $\chi$PT for the 
 neutral pion photoproduction from nucleons:   
\begin{eqnarray}
E^{tree}_{0^+}\left(\gamma p \rightarrow \pi^0 p \right) =& 
\frac{-e }{8 \pi f_\pi} (D+F)&\left[ \mu_\pi - 
 \mu^2_\pi\frac{3+\kappa_p}{2}\right]
\nonumber\\
E^{tree}_{0^+}\left(\gamma n \rightarrow \pi^0 n \right) =& 
\frac{-e }{8 \pi f_\pi}(D+F)&\left[ \hspace{8.0mm}  
\mu^2_\pi\frac{\kappa_n}{2}\hspace{4.5mm}\right].
\label{trees}
\end{eqnarray}
In the above, $\mu_\pi \equiv m_\pi/M_N$, 
  $D+F = g_A = 1.26$ and $\kappa_p = 1.79$ and $\kappa_n = -1.91$ are the anomalous 
magnetic moments of the proton and neutron, respectively.
 
\section{One loop, $O(p^2)$ results}
The six nonvanishing, $O(p^2)$ 1--loop graphs for neutral pion 
photoproduction from a proton is shown in Fig. (2).   
We note that the   graphs resulting from the higher order expansion 
of the axial current ({\it i.e.} the three pion term of $A^\mu$) vanish due to 
$SU_f(3)$ group factors. These graphs have not been exhibited. For a
similar reason, explained later, the analogs of graphs of Figs. (2c)
and (2d) with charged pion in the loop have not been exhibited.     In Fig. (2) the
sum of each graph and it's 
corresponding crossed graph is ultraviolet finite. Working in the
nucleon's rest frame   suitably choosing the loop momentum $k$ in each
of the two graphs one obtains expressions identical in all respects
except the baryon propagator which appears as  form $1/(\pm v\cdot
k+i\eta)$.  Hence, only the $i\pi\delta(v\cdot k)$ term contributes in
the sum of the two graphs yielding ultraviolet finite results.  
 Picking up the 
baryon poles and then using dimensional regularization\cite{us1} 
 for evaluating the remaining three Euclidean space integrals allows for 
the simplest evaluation of the graphs (taken pairwise).  For example, the 
two graphs in Fig. (2c) and Fig. (2d) give the following contribution to 
the transition amplitude $T$, Eq.~(\ref{Mdef}):
\begin{eqnarray}
&&{\cal M} ^{K K\, loop}_{0^+}(\gamma p \rightarrow \pi^0 p ) \hfill\nonumber\\
&=& 
\frac{-\imath e F}{f_\pi^3}\bar{H} 
\int \frac{d^4 k}{(2 \pi)^3} 
\frac{\delta (v \cdot k) \, v \cdot (r + q) \, S \cdot k \, \epsilon \cdot k}
{(k^2 - m^2_K) ((q+k)^2 - m^2_K)} \, H 
\nonumber\\
&=& \frac{-\imath e F m_\pi}{4 \pi^3 f_\pi^3}\bar{H} 
\int d^3 k
\frac{\bar{S} \cdot \bar{k} \, \bar{\epsilon} \cdot \bar{k}}
{(\bar{k}^2 + m^2_K) (\bar{k}^2 + 2 \bar{q} \cdot \bar{k} + m^2_K)} \, H 
\nonumber\\
&=& \frac{-\imath e F m_\pi}{8 \pi^3 f_\pi^3} 
\bar{H} \bar{S} \cdot \bar{\epsilon} \, H \pi^{3/2} \Gamma(-\frac12) 
\int_0^1 dx \sqrt{m_k^2 - x^2 m_\pi^2} 
\nonumber\\
&=& - \imath \bar{H} S \cdot \epsilon \, H 
\frac{ e F }{8 \pi f_\pi^3}m_\pi \nonumber\\
&&\hspace{.5in}\times 
\left[ \sqrt{m_K^2 - m_\pi^2} + \frac{m_K^2}{m_\pi} sin^{-1}
\left(  \frac{m_\pi}{m_K} \right) \right], \label{KK}
\end{eqnarray}
yielding a contribution to the amplitude $E_{0^+}$ of:
\begin{eqnarray}
E^{K K\, loop}_{0^+}&&(\gamma p \rightarrow \pi^0 p ) = 
 \frac{ e F }{64 \pi^2 f_\pi^3}m_\pi \hfill\nonumber\\
&&\times \left[ \sqrt{m_K^2 - m_\pi^2} + \frac{m_K^2}{m_\pi} sin^{-1}
\left(  \frac{m_\pi}{m_K} \right) \right].
\end{eqnarray}
The contribution\cite{Bernardetal} of Figs. (2a) and (2b) can be obtained 
from the above by making the obvious substitution $m_K \rightarrow m_\pi$ and  
also replacing $F$ by $D + F$ to account for the different SU$_f$(3) couplings.

 Proceeding similarly,   the two graphs in Figs. (2e) and (2f) give the contribution
\begin{eqnarray}
&&{\cal M} ^{K\, loop}_{0^+}(\gamma p \rightarrow \pi^0 p )  \hfill\nonumber\\
&=& 
 \imath \bar{H} S \cdot \epsilon  \, H \frac{e F}{2 f_\pi^3}
\int \frac{d^4 k}{(2 \pi)^3} 
\frac{\delta (v \cdot (r+k) )\, v \cdot (r - k)}{k^2 - m^2_K} 
\nonumber\\
&=& \imath \bar{H} S \cdot \epsilon  \, H
 \frac{e F}{4 \pi f_\pi^3} m_\pi  \sqrt{m_K^2 - m_\pi^2}. \label{K}
\end{eqnarray}
From this expression it is obvious that the summed contribution from the 
analogous two graphs with a charged pion propagating in the loops is zero.  

Summing the two classes of kaon loop contributions gven by
Eqs.~(\ref{KK}) and (\ref{K}) we obtain the following contribution to
the $E_{0^+}$ amplitude: 
\begin{eqnarray}
E^{K K\, loop}_{0^+} (\gamma p \rightarrow \pi^0 p ) &+&E^{K \,
loop}_{0^+}(\gamma p \rightarrow \pi^0 p )= \nonumber \\
 &&\frac{ e F }{64 \pi^2 f_\pi^3}m_\pi  
 \times  f(m_K, m_\pi),
\end{eqnarray}
where the function $f(m_K, m_\pi)$ is given by 
\begin{equation}
f(m_K,m_\pi) = 
\frac{m_K^2}{m_\pi} sin^{-1}\left(\frac{m_\pi}{m_K}\right) - \sqrt{m_K^2 - m_\pi^2}.
\end{equation}
This results agrees with the result reported by Steininger {\it et
al.}\cite{SandM}.\footnote{An earlier version of the present  paper had
an error which was pointed out in Ref.~\cite{SandM}.}
  
Combining these results with the tree--level contributions of
Eq.~(\ref{trees}), one 
obtains that the complete $O(p^2)$ expression for threshold pion photoproduction 
from nucleons in $SU_f(3)$ $\chi$PT is:
\begin{eqnarray}
&&E_{0^+}\left(\gamma p \rightarrow \pi^0 p \right) = 
-\frac{e g_A}{8 \pi f_\pi} \hfill \nonumber\\
&&\times \left[ \mu_\pi - \mu^2_\pi\frac{3+\kappa_p}{2}
- \frac{m_\pi^2}{16 f_\pi^2} - \frac{F}{g_A}
 \frac{m_\pi f(m_K, m_\pi)}{8 \pi f^2_\pi} \right] 
\label{ampsp}\\
&&E_{0^+}\left(\gamma n \rightarrow \pi^0 n \right) = 
-\frac{e g_A}{8 \pi f_\pi} \hfill\nonumber\\
&&\times \left[ \hspace{8.0mm}  \mu^2_\pi\frac{\kappa_n}{2}
- \frac{m_\pi^2}{16 f_\pi^2} - \frac{F-D}{g_A} 
\frac{m_\pi f(m_K, m_\pi)}{16 \pi f^2_\pi} \right].
\label{ampsn}
\end{eqnarray}

\section{Discussion}

According to a  recent analysis\cite{latestBernard} of the world's
data\cite{pidata}  
 the current best value of $E_{0^+}\left(\gamma p
\rightarrow \pi^0 p \right)$ is $\sim -1.5 \times10^{-3}m_\pi^{-1}$.  
The numerical values of the various contribution to the photoproduction
amplitudes are listed below.  The kaon loop contributions are
calculated  by  using $F=0.4$ and $D=0.86$.  The values are listed in
the same order of appearance as that of the terms in Eqs~(\ref{ampsp})
and (\ref{ampsn}).
\begin{eqnarray}
E_{0^+}\left(\gamma p \rightarrow \pi^0 n \right)
&=&(-3.28+1.13+3.01+0.11) \nonumber \\
&&\times10^{-3}m_\pi^{-1} \label{valuep} \\
E_{0^+}\left(\gamma n \rightarrow \pi^0 n \right) &=&(0+.45+3.01-0.07)\nonumber \\
&&\times10^{-3}m_\pi^{-1} \label{valuen} \end{eqnarray} As mentioned
earlier the total tree level result for $E_{0^+}\left(\gamma p
\rightarrow \pi^0 n \right)$ is reasonably close to the experimental
value. The one-loop result involving pion is of the same size as the
tree level result and upon adding it one gets an amplitude with wrong
sign. The kaon loop contribution is negligible and consequently does
not help matters.  There is distinct indication that $\chi$PT is not
converging in the present situation.

It has been suggested~\cite{latestBernard}    that in     SU$_f(2)$   
theory convergence 
sets in at ${\cal O}(p^3)$ or at an even higher order.  The same suggestion 
 presumably applies  to the SU$_f(3)$  theory too.  
Of course, in both cases one encounters  new,
unknown counterterms.   One is likely to need the threshold
photoproduction data themselves to fix  these terms.   Thus one loses
predictive power.   

We suggest   that the {\it difference} between the proton and 
neutron $\pi^0$ photoproduction amplitudes would  provide a 
  test of the convergence properties of chiral perturbation theory.  The
SU$_f$(3) result for the difference at ${\cal O}(p^2)$ level is
 \begin{eqnarray}
&&E_{0^+}\left(\gamma p \rightarrow \pi^0 p \right) - 
E_{0^+}\left(\gamma n \rightarrow \pi^0 n \right) \left |_{SU_f(3)}
\right. \nonumber\\
&&\approx -\frac{e g_A}{8 \pi f_\pi} 
 \left[ \mu_\pi - \mu^2_\pi\frac{3+\kappa_p+ \kappa_n}{2} - 
\frac{f(m_K, m_\pi) m_\pi}{16 \pi f_\pi^2}\right] \nonumber\\
&&(-3.28+0.68+0+0.17)\times 10^{-3} /m_{\pi^+}\nonumber\\
&&\approx -2.5 \times 10^{-3} /m_{\pi^+}. \label{SUf3dif}
\end{eqnarray} The last but one line gives the breakdown according to
the various contributors. We see that the value at ${\cal O}(p^1)$ is
$-3.3\times 10^{-3} /m_{\pi^+}$ and the total contribution of the
${\cal O}(p^2)$ is $0.8\times 10^{-3} /m_{\pi^+}$. There is a distinct
tendency towards convergence. It is   legitimate to hope that the higher
chiral power terms, which are suppose to cancel out the one-loop $O(p^2)$
terms in the individual amplitudes, do not contribute much to the
difference.   Then comparing the predicted value of
\mbox{Eq.~(\ref{SUf3dif})} with   experiment  will constitute a  test
of $\chi$PT.   If the amplitude is truly convergent the numerical value
of $\approx -2.5 \times 10^{-3} /m_{\pi^+}$ should compare well with
the experimental data, something which cannot be claimed for the
results of Eqs~(\ref{valuep}) and (\ref{valuen}).

\section{Conclusions}

It is known that the tree level result at ${\cal O}(p^2)$   of
SU$_f(2)$ chiral perturbation theory result for the
$E_{0^+}\left(\gamma p \rightarrow \pi^0 p \right)$ amplitudes is in
fair agreement with the experimental result. But addition of one-loop 
${\cal O}(p^2)$  contribution spoils the agreement radically, to the
extent of changing the sign of the amplitude. Since kaon and pion loop
contributions tend to cancel each other in magnetic moment calculation,
we examined if a similar situation occurs in neutral pion
photoproduction amplitude.  We find that contrary to the expectation,
extension from SU$_f(2)$ to SU$_f(3)$ theory makes very little change.
In fact, it  makes the disagreement 
with the data marginally worse.

The radical change produced by the one-loop  ${\cal O}(p^2)$ 
contribution   raises questions about
convergence of the theories.  One may speculate that convergence sets
in at higher chiral
power. We find that this speculation leads to a testable 
prediction for  the   amplitude difference, $E_{0^+}\left(\gamma p \rightarrow
\pi^0 p \right) - 
E_{0^+}\left(\gamma n \rightarrow \pi^0 n \right)$.
It is reasonable to expect a convergent result  
at the ${\cal O}(p^2)$ level.  Specifically, the prediction for the
difference amplitude is $\sim -2.5 \times 10^{-3} /m_{\pi^+}$. An
experimental test of this prediction will shed light on the convergence issue.

This discussion makes clear that it is important to obtain experimental
data on  $\pi^0$
off neutron. Such experiments are being considered at CEBAF.

\bigskip
{\centerline{ACKNOWLEDGEMENTS}} 
We thank M. Birse for getting us interested in this problem and 
T. D. Cohen and C. Bennhold for discussions.
We also thank N. Kaiser for useful correspondence.
This work was supported in part by DOE Grant DOE-FG02-93ER-40762.

\newpage
\begin{figure}
\vglue 2.5in 
\caption{The tree--level graphs that contribute to neutral pion photoproduction 
through $O(p^2)$ in the relativistic theory.  These are reduced to seagull terms in 
the heavy baryon formulation.}
\end{figure}

\begin{figure}
\vglue 4.5in
\caption{The six lowest order, one--loop contributions to neutral pion 
photoproduction from a proton  in $SU_f(3)$  $\chi$PT. }
\end{figure}

\begin{references}
\bibitem{andGasser}
V. Bernard,  J. Gasser, N. Kaiser and U.-G.  Mei{\ss}ner, 
Phys. Lett. {\bf B268}, 291 (1991);
V. Bernard,  N. Kaiser and U.-G.  Mei{\ss}ner, 
Nucl. Phys. {\bf B383}, 442 (1991).

\bibitem{pidata}
E. Mazzucatao {\it et al.}, 
Phys. Rev. Lett. {\bf 57}, 3144 (1986);  
R. Beck {\it et al.}, 
Phys. Rev. Lett. {\bf 65}, 1841 (1990); 
T. P. Welch {\it et al.},  
Phys. Rev. Lett. {\bf 69}, 2761 (1992).

\bibitem{AaronBarry}
For a review of the pre-one-loop analysis of $\chi$PT, see 
A. M. Bernstein and B. R. Holstein, 
Comments Nucl. Part. Phys. {\bf 20}, 197 (1991).

\bibitem{latestBernard}
V. Bernard, N. Kaiser, and U.-G.  Mei{\ss}ner, 
{\it Neutral pion photoproduction off Nucleons Revisited}, 
 hep-ph/9411287,  Strasbourg preprint CRN 94-62, 
Bonn preprint TK 94 18, (1994);  to appear in Z. Phys. C.

\bibitem{gangof4}
E. Jenkins, M. Luke, A. V. Manohar, and M. J. Savage,
Phys. Lett. {\bf B302}, 482 (1993).

\bibitem{us2}
M. K. Banerjee and J. Milana, 
{\it The Decuplet Revisited in $\chi$PT}, 
University of Maryland preprint, DOE/ER/40762--065, 
UMPP \#96--19 (1995).

\bibitem{known1}
T. Mart, C. Bennhold, and C. E. Hyde--Wright, 
Phys. Rev. {\bf C51}, R1074 (1995); 
Zh. Li, 
Phys. Rev. {\bf C52}, 1648 (1995); 
J. C. David, C. Fayard, G. H. Lamot, and B. Saghai, 
{\it Electromagnetic Production of Associated Strangeness}, 
DAPNIA preprint, DAPNIA--SPhN-95-52 (1995).

\bibitem{known2}
M. Benmerrouche and N. C. Mukhopadhyay, 
Phys. Ref. Lett. {\bf 67}, 1070 (1991); 
M. Benmerrouche, N. C. Mukhopadhyay and J.F. Zhang, 
 Phys. Rev. {\bf D51}, 3237 (1995); 
C. Bennhold and H. Tanabe, 
Nucl. Phys. {\bf 530A}, 625 (1991); 
L. Tiator, C. Bennhold, and S. S. Kamalov, 
Nucl. Phys. {\bf 580A}, 455 (1994).

\bibitem{Bernardetal}
V. Bernard, N. Kaiser, J. Kambor and U.-G.  Mei{\ss}ner,
Nucl. Phys. {\bf B388}, 315 (1992).

\bibitem{JenkMan1}
E. Jenkins, and A. V. Manohar, 
Phys. Lett. {\bf B255}, 558 (1991).

\bibitem{pedagogy}
For a lucid introduction to the heavy quark effective theory, as
well as to Chiral Perturbation theory, see, J. F. Donoghue, E.
Golowich, and B. Holstein, ``Dynamics of the Standard Model'',
(Cambridge University Press, Cambridge England, 1992).

\bibitem{ColeGlas}
S. Coleman and S. L. Glashow,  
Phys. Rev. Lett. {\bf 6}, 423 (1961).
S. Weinberg, Phys. Lett. {\bf B251}, 288 (1990).

\bibitem{GassNuc}
J. Gasser, M. E. Sainio, and A. Svarc, Nucl. Phys. {\bf B307},
779 (1988).

\bibitem{Wcount} 
S. Weinberg, Phys. Lett. {\bf B251}, 288 (1990).
Wherein in the heavy baryon or nonrelativistic version of the theory, 
the chiral power $\nu$ of a one--nucleon 
irreducible graph with $L$ loops, $V_i$ vertices of type 
$i$ characterized by $d_i$ derivatives or
factors of $m$ and $n_i$ number of nucleon fields is given by 
$\nu = 2 - N + 2L + \Sigma_i V_i (d_i  - 1)$
 
\bibitem{us1}
M. K. Banerjee and J. Milana,
Phys. Rev. {\bf D52}, 6451 (1995).

\bibitem{SandM}
S. Steininger and U.-G.  Mei{\ss}ner, 
University Bonn preprint TK 96 04, (1996).

\end{references}
\end{document}